\documentstyle[prl,aps,epsf]{revtex}

\begin{document}

\begin{center}

{\Large \bf Large non-adiabatic hole polarons and matrix element effects in the
angle-resolved photoemission spectroscopy of dielectric cuprates}

\vspace{0.5cm}

A.S. Moskvin,  E.N. Kondrashov, V.I. Cherepanov

\vspace{0.5cm} Department of Theoretical Physics, Ural State University 620083,
Ekaterinburg, Russia

\vspace{0.5cm}

\end{center}

It has been made an extention of the conventional theory based on the
assumption of the well isolated  Zhang-Rice singlet to be a first
electron-removal state in dielectric copper oxide. One assumes the photohole
has been localised on either small (pseudo)Jahn-Teller polaron or large
non-adiabatic polaron enclosed one or four to five $CuO_4$ centers,
respectively, with active one-center valent $(^{1}A_{1g}-{}^{1,3}E_{u})$
manifold. In the framework of the cluster model we have performed a model
microscopic calculation of the ${\bf k}$-dependence of the matrix element
effects  and photon polarization effects for the angle-resolved photoemission
in dielectric cuprate like $Sr_{2}CuO_{2}Cl_{2}$. We show that effects like the
''remnant Fermi surface'' detected in ARPES experiment for
$Ca_{2}CuO_{2}Cl_{2}$ may be, in fact, a reflection of the matrix element
effects, not a reflection of the original band-structure Fermi surface, or the
strong antiferromagnetic correlations. The measured dispersion-like features in
the low-energy part of the ARPES spectra may be a manifestation of the complex
momentum-dependent spectral line-shape of the  large PJT polaron response, not
the dispersion of the well-isolated Zhang-Rice singlet in antiferromagnetic
matrix.



\vspace{0.4cm}
 {\large \bf Introduction}
\vspace{0.2cm}

Angle-resolved photoemission spectroscopy  (ARPES) is considered
to be a key experiment to elucidate a number of principal issues
of electronic theory related to the unconventional properties of
cuprates, manganites, nickellates, bismuthates, and other strongly
correlated oxides. These are the quasiparticle dispersion, the
Fermi surface, (pseudo)gap behaviour and other ones. Many researchers
consider namely ARPES data to be a main evidence in favour of
either model of electronic structure and energy spectrum.

However, the current practice of ARPES activity in oxides does not justify
these expectations. Up to now the  ARPES measurements are treated as a rule in
the framework of oversimplified ''three-stage'' approach. The first
''experimental'' one relates straightforwardly to  measurements  of the
($E,{\bf k}$)-dependencies of photocurrent intensity. Any problems here are
related to a sample preparation,  experimental setup, and attempts to avoid
such external parasitic phenomena as a charging effect. The very ''dangerous''
second ''experimental-theoretical'' stage implies the preliminary treatment and
often incorporates a ''hidden'' interpretation based on a number of seemingly
inviolable statements like:

1. ARPES data in a wide spectral range could be self-consistently described by
the conventional band models.  Electronic structure and energy spectrum near
the Fermi level are formed by Landau quasiparticles with Fermi statistics which
fill an energy band.

2. The peaks position for the photocurrent intensity links to the peaks in the
quasiparticle spectral density. In other words, ARPES essentially measures the
one-particle spectral function of the initial state.

3. Some authors go further and  make use of the standard method which provides
the information on the Fermi surface from ARPES data in the conventional band
metals. First, one find the occupation probability, $n({\bf k})$, by
integrating the ARPES spectral function $A({\bf k},E)$ over energy.
Experimentally, one choses an energy window for integration, thus becoming the
relative $n({\bf k})$. Then, the drop of  the relative $n({\bf k})$ is used to
determine the Fermi surface. In practice, one defines $k_F$ as the locus of
points of maximum gradient of $n_{{\vec k}}$ or  as the point of steepest
descent in the relative $n({\bf k})$. Naturally, this method implies the simple
metallic-like electronic structure, where the identification of a Fermi surface
is convincing.

In our opinion,  this ''band-like'' paradigm supported only by the simplest
model like that of free electrons, turns out to be at least questionable in all
points in the case of strongly correlated oxides. Firstly, an ability of band
models, even modified like  LDA+U \cite{Anisimov}, to yield a  relevant
description of electron spectra for strongly correlated oxides is merely a
misleading, at present there are not convincing examples for such a
description, in particular, for the lowest-lying occupied band, or the
so-called first electron-removal states. The typical errors in locating the
$final$ $states$ for $ab$ $initio$ band structures are generally expected to be
of the order of a few eV's (!). Large number of experimental data evidences in
favour of unconventional ''non-Landau '' nature of quasiparticles and
occurrence of unusual correlations in cuprates and other oxides. Traditional
interpretation of ARPES data like a simple accordance between the
photointensity peak position and the quasiparticle spectral density peak
implies a full neglect of several factors each of which is capable to result in
crucial reconsideration of the ARPES data: i) the matrix element effects, or
the intensity dispersion; ii) surface effects; iii) effects of multiple
scattering; iv) effects of finite lifetimes of the initial and final states; v)
effects of coupling with phonons, spin excitations and other possible degrees
of freedom; vi) effects of configurational interaction; vii) effects of
electron inhomogeneity, for instance, in doped cuprates. For insulating samples
it is difficult to escape the charging effects.

Importance of matrix element effects and the ARPES intensity dispersion for
different cuprates was underlined earlier
\cite{Alexandrov,Eroles,Bansil,Haffner1,Haffner2,Seibel}, and a role played by
configurational interaction was partially illustrated in Ref.
\cite{Ovchinnikov} for $Sr_{2}CuO_{2}Cl_{2}$. The hole polaron formation due to
electron-phonon coupling and its manifestation in ARPES  was demonstrated by
Alexandrov and Dent \cite{Alexandrov} for cuprates like Y123 and Y124. One
should be noted that all the above mentioned citations relate to the
nontraditional approach in ARPES.

The third, purely theoretical stage implies a qualitative and
quantitative description of the dispersion law and other
quasiparticle properties obtained on the foregoing stage and to be
considered as ''experimental data'' (?!). Unfortunately, too often
this stage looks like a formal fitting with either model theory, and
with no critical inspection of experimental data.
Here one might mention, for instance, numerous papers with
theoretical treatment of ''experimental'' ARPES data for
$Sr_{2}CuO_{2}Cl_{2}$ in the framework of various extended  $t-J$,
$t-t'-J$, $t-t'-t''-J$ models \cite{Wells,Nazarenko}.

Summarizing, we see  this ''three-stage'' approach could result in
a natural doubts as to conclusions based on such an oversimplified
interpretation.

As a convenient demonstrative model system for ARPES in the insulating layered
copper oxides one might be chosen  oxychlorides  $Sr_{2}CuO_{2}Cl_{2}$ and
$Ca_{2}CuO_{2}Cl_{2}$, which are isostructural to famous 214 system
$La_{2}CuO_{4}$.  The oxychloride $Sr_{2}CuO_{2}Cl_{2}$ is one of the most
popular model system for insulating phase of the high-$T_c$ cuprates and is
intensively studied both experimentally and theoretically.

In this tetragonal  antiferromagnetic with nearly ideal $CuO_2$ planes there
are chlorine atoms instead of apex oxygens with considerably larger
$Cu-Cl_{apex}$ separation  (2.86\AA) than that of $Cu-O_{apex}$ (2.42\AA) in
$La_2CuO_4$. Hence, in $Sr_2CuO_2Cl_2$ one appears a real opportunity to
examine the $CuO_2$ plane states, both copper and oxygen, without ''parasitic''
contribution of apex oxygens. At present there are a rather large number of
experimental data for $Sr_2CuO_2Cl_2$ obtained with the help of optical
spectroscopy  \cite{optics}, X-ray photoemission (XPS) \cite{x-ray},
ultraviolet photoemission (UPS) \cite{Fujimori}, X-ray absorption spectroscopy
(XAS) \cite{XAS,XAS1}, electron energy loss spectroscopy (EELS)
\cite{Wang,EELS,EELS1}, angle-resolved photoemission spectroscopy
\cite{Haffner1,Haffner2,Wells,LaRosa,Kim}. Similar experimental ARPES results
were obtained recently by F. Ronning et al. \cite{Ronning} for $Ca_2CuO_2Cl_2$
which is a full analog of $Sr_2CuO_2Cl_2$.

The authors \cite{Wells} state: ''This is a measurement of the dispersion of a
single hole in an antiferromagnetic background, a problem that has been heavily
investigated theoretically''. It seems, this is a very strong and unambiguous
statement, which oversimplifies the problem and does not account of many
alternative scenarios. Moreover, the experimental data \cite{Wells} could
hardly considered to be a reliable basis set for any decisive conclusions.
Indeed, the spectral dependence of the photocurrent intensity shows a complex
nature of the low-energy photo-hole states with a strong and dispersive
contribution of the high-lying states.

It is very important to emphasize that, while the ARPES spectra in various
oxides displayed as the photointensity against $E,{\bf k}$ curves show the
dispersion-like features (see, e.g. Fig.1, Ref.\cite{Wells}, or Fig.3,
Ref.\cite{Ronning}), the ''dispersing'' states which ''peak positions'' are
plotted are extremely broad, with width comparable to binding energy, and these
simply cannot be thought of as quasiparticles. This general point is true at
all ${\bf k}$'s. In addition, in the most part of the Brillouin zone there are
no well-defined ARPES peaks at all. Moreover, the observed peak position may
rather strongly depend on the energy resolution.

The similar to $Sr_2CuO_2Cl_2$ experimental results  were obtained
by F. Ronning et al. \cite{Ronning} for $Ca_2CuO_2Cl_2$. The improved spectral quality
allows  authors to reveal a steep drop in spectral intensity across a contour that
is close to the Fermi surface predicted by the band calculation. They concluded that
the Fermi surface, which is destroyed by the strong Coulomb interactions, left a remnant
in this insulator with a volume and shape similar to what one expects if the strong
electron correlation in this system is turned off. The lowest energy peak exhibits
a dispersion with approximately the $|cosk_xa-cosk_ya|$ form along this remnant Fermi surface,
 in other words the strong correlation effect deforms this otherwise iso-energetic contour
 (the non-interacting Fermi surface) into the form that resembles the d-wave like pseudogap
 dispersion  with  a very high energy scale of 320 meV \cite{Ronning}.
 The authors \cite{Ronning} consider the d-wave like dispersion of the parent insulator
 to be the underlying reason for the pseudogap in the underdoped cuprates.
They follow straightforwardly the simplest model approach to ARPES and
electronic structure of the photo-hole, and  do not consider either alternative
approaches, albeit they   mention some ''parasitic'' effects of matrix elements
and photon polarization, especially, if these effects hardly keep within their
generic model. The approach \cite{Ronning} is demonstrative for many papers on
ARPES, in particular, addressing the Fermi surface problem.

At present there are several papers with experimental ARPES data on
$Sr_{2}CuO_{2}Cl_{2}$ and $Ca_{2}CuO_{2}Cl_{2}$ systems
\cite{Haffner1,Haffner2,Wells,LaRosa,Kim,Ronning} distinguished by different
energy resolution from 25 $meV$ \cite{LaRosa} to 105-115 $meV$
\cite{Haffner1,Haffner2}, different photon energy from 10 $eV$ to 80 $eV$
\cite{Haffner1,Haffner2,Durr}, light polarization \cite{LaRosa,Durr}, and the
measurement temperature. Despite the many common features there are some
important departures. So, contrary to findings \cite{Wells}, the strong
''quasiparticle'' dispersion was observed in $Sr_{2}CuO_{2}Cl_{2}$ for the
$(\pi ,0)$ direction with strong and unusual dependence of the peak amplitude
on the light polarization \cite{LaRosa}. Kim {\it et al.} \cite{Kim} performed
the ARPES measurements in Sr-oxychloride  at 150 K, well below the N\'eel
temperature, and have confirmed that the spectrum in the $(\pi ,\pi )$
direction consists of a single relatively sharp peak near $(\pi /2,\pi /2)$.
However, the spectra along $(\pi ,0)$ direction are very broad and consist of
at least two peaks separated by about 0.4 $eV$. Just recently, using the ARPES
spectra of $Sr_{2}CuO_{2}Cl_{2}$ as an example, the authors
\cite{Haffner1,Haffner2} have experimentally demonstrated a significant impact
of electron-photon matrix elements on both the relative spectral intensity and
the shape of a low-energy feature in ARPES spectrum.

Many experimental ARPES data and conclusions on $Sr_{2}CuO_{2}Cl_{2}$ known up
to now were revised very recently in a detailed investigation by C. D\"urr {\it
et al. } \cite{Durr}. The authors observed the marked oscillating photon-energy
dependence of the photoemission signal of the first electron-removal state,
which was attributed to the diffraction of the photoelectron wave on the
$c$-axis periodically arrayed $CuO_2$ planes. They found a strong polarization
dependence in ARPES spectra along the high-symmetry directions, compatible with
that expected for a Zhang-Rice singlet. The ratio between the coherent and
incoherent spectral weight in ARPES spectra near the first electron-removal
state appears to be photon-energy dependent. Among several mechanisms of this
puzzling effect the authors \cite{Durr} mention the possible contribution of
the low-lying electronic states other than the Zhang-Rice singlet.

One of the remarkable spectral features clearly revealed in the ARPES measurements
 for both insulating cuprates \cite{Wells,Ronning} is a significant
 low-energy spectral weight in the $\Gamma$ (0,0) point that accordingly to
 symmetry of the electric-dipole matrix elements   implies
 a significant weight of the purely oxygen odd $e_u$-symmetry photo-hole state.
 However, in the framework of traditional approach one prefers to take no notice
  of this feature.

Our paper has not for an object the elaboration of the general theory of ARPES
in the strongly correlated oxides. The authors would like to emphasize on the
possibly simple but real examples an importance of some factors such as matrix element effects, formation of the hole non-adiabatic polaron, often to be
disregarded in the framework of conventional approaches to interpretation of
ARPES data in copper oxides. At the same time the paper could be considered as
a first step in  elaboration of an original theory of ARPES in the strongly
correlated oxides.

The paper is organised as follows. In Section I we present a brief summary of
results of quantum chemical modelling the electronic structure and energy
spectrum for the $CuO_{4}^{6-}$ and $CuO_{4}^{5-}$ centers with one and two
holes, respectively. One considers the formation of  small and large
non-adiabatic polarons.
 It should be mentioned here that the concept of polarons led to the discovery of
 the copper oxide superconductors. Section II incorporates an analysis of the generic
  expression for
the photoemission current intensity in the framework of the PJT polaronic
approach. In Section III we present a model calculation of dipole matrix
elements which describe a transition of a bound $\gamma\mu$ electron to unbound
state. An illustrative modelling of the $k$-, and polarization dependence for
the matrix element part of the ARPES intensity  is given in Section IV.

\vspace{0.4cm}
{\large \bf 1. Electronic structure of the hole state in cuprates. From small to large PJT polaron}
\vspace{0.2cm}

One of the peculiar properties of the doped cuprates is a cross-over from localized to itinerant electronic behaviour with an inhomogeneous distribution of electronic states which description in a wide range of compositions from  a single-valent antiferromagnetic insulator to a mixed valent bad metal is a fundamental problem of the solid state physics.

Unfortunately, at present there is no a general consensus concerning the nature of valent electron and hole states which form optical and electron (PES, XPS, EELS) spectra even in insulating cuprates like  $La_2CuO_4$, $YBa_2Cu_3O_6$, $Sr_2CuO_2Cl_2$.
It is becoming increasingly difficult to reconcile experimental results with the expectations of the simple band models and Fermi liquid theory.

One of  the firmly established facts relates the $CuO_2$ plane  character of the valent states  with the  $b_{1g}(d_{x2-y2})$ hole ground  state for the $CuO_{4}^{6-}$ center.  Various spectroscopic methods link the low-energy excitations in the range $E\geq 2$ eV to the charge transfer $O2p\rightarrow Cu3d$ within the  $CuO_2$ plane. At a first glance, optical and electron spectra manifest both
localised and delocalised, or band character of electronic states. Lack of reliable interpretation of optical and electron spectra in strongly correlated oxides appears to be a result of difficulties in theoretical description of strong covalency and strong intracenter and intercenter correlations.

The parent cuprates such as $CuO$,  $Sr_2CuO_2Cl_2$, $La_2CuO_4$, $YBa_2Cu_3O_6$ provide typical examples of systems with strong local correlations when dielectric antiferromagnetic ground state is mainly specified by potential energy of electron-electron coupling. Kinetic energy would prefer formation of half-filled band with typically metallic behavior. Occurrence of the dielectric antiferromagnetic phase of the parent cuprates itself and wide opportunities of the interacting $CuO_4$-centers model in explanation of many optical and electron spectra for various copper oxides could be considered to be a convincing evidence in favor of quasiparticle states localised predominantly on the $CuO_4$-clusters which do not obey conventional band model description. This is a common place as for  $Cu3d-O2p$-hybrid states, but purely oxygen non-bonding states are currently described as the band ones. Such a discrimination is partially based on a rather simplified relation between correlation and Slater electrostatic integrals like $F_{0}(3d3d)>F_{0}(2p2p)$, and typical, for instance, for the Anderson impurity model.

However, correlation effects appear to be of particular importance
for oxygen hole states which was pointed out by Hirsch et al.
\cite{Hirsch} in their theory of ''anionic metal''. A simple model
for localisation of the oxygen holes and some other puzzling
consequences of conjectured \cite{Hirsch} multiplet structure for
anionic background were considered recently \cite{background}.  A
strongly correlated behaviour of the $O2p$-holes  to some extent
gives them equal rights with $3d$-holes and leads to
reconsideration of many conventional approaches and models which
ignore strong $O2p$-correlations, for instance, the Anderson
impurity model which considers purely oxygen states to be
band-like.

So, a quantum-chemical cluster approach appears to be more relevant for  description of strongly correlated oxides as it allows to account for correlation effects both for cation and anion states by the most optimal manner. It seems, the parent copper oxides can be treated by a conventional quantum-chemical ligand-field theory, so a localized description can be considered to be a relevant starting point for modelling the electronic structure both in ground and excited states.

 First of all, the modelling implies a choice of the effective basis set of atomic orbitals, such as $Cu3d$- and $O2p$-orbitals in the cuprates. Then we  introduce the $CuO_4$ center being the minimal atomic cluster that posesses the relevant point symmetry $D_{4h}$ and incorporates the main $Cu3d-O2p$ covalent bond. This center could be considered as an elementary building block to effectively model the electronic structure both in ground and excited states. Along with the symmetry and covalency the $CuO_4$ center allows to account for the electronic and vibronic correlations.

\vspace{0.4cm}
{\bf 1.1. Electronic structure and energy spectrum for the $CuO_{4}^{6-}$ and $CuO_{4}^{5-}$ centers in cuprates}
\vspace{0.2cm}

Slightly distorted tetragonal $CuO_4$ center is the only common cell of
crystalline and electronic structure in a wide variety of  high-$T_c$ copper
oxides  and related parent systems. Restricted atomic basis of five  $Cu3d$ and
twelve $O2p$ atomic orbitals for the  $CuO_4$ cluster with $D_{4h}$ symmetry
yields seventeen symmetrised orbitals with $a_{1g}$, $a_{2g}$, $b_{1g}$,
 $b_{2g}$, $e_{g}$ (even - gerade) and
$a_{2u}$, $b_{2u}$, $e_{u}$ (odd - ungerade) symmetry. The even $Cu3d$ orbitals
with $a_{1g}(3d_{z^2})$, $b_{1g}(3d_{x^2-y^2})$,
 $b_{2g}(3d_{xy})$, $e_{g}(3d_{xz}$, $3d_{yz})$ symmetry hybridize with even
$O2p$ orbitals with the same symmetry thus forming the appropriate bonding  $\gamma ^{b}$ and   antibonding  $\gamma ^{a}$ molecular orbitals.
 Among the odd orbitals only  $e_{u}(\sigma)$ and $e_{u}(\pi)$ ones  hybridize each other due to a strong  $O2p-O2p$ coupling to form bonding  and antibonding purely oxygen molecular orbitals $e_{u}^{b}$ and $e_{u}^{a}$, respectively. Purely oxygen $a_{2g},a_{2u},b_{2u}$ orbitals are nonbonding. All the plane molecular orbitals could be subdivided to
$\sigma$ ($a_{1g},b_{1g},e_{u}(\sigma)$) and  $\pi$
($a_{2g},b_{2g},e_{u}(\pi)$) ones, depending on the orientation of $O2p$
orbitals.

Quantum-chemical cluster approach allows the optimal account for electrostatic
correlations and description of a rather complex electronic structure of the
two-hole  $CuO^{5-}_{4}$ center. Firstly, one should note an importance of
configurational interaction, which could be rather simply illustrated by the
model cluster calculation \cite{Moskvin94}. Indeed, the wave function and
energy of the ground state term $^{1}A_{1g}$ (Zhang-Rice singlet) exhibits an
essential coupling for three configurations like   $b_{1g}b_{1g}$:
$$
|^{1}A_{1g}>= 0.82|(b_{1g}^{b})^{2}> + 0.55 |b_{1g}^{a}b_{1g}^{b}> - 0.16
|(b_{1g}^{a})^{2}>.
$$
If to consider the   $^{1}A_{1g}$ singlet originating from the only   $(b_{1g}^{b})^{2}$ configuration and to represent its energy as follows
$$
E(^{1}A_{1g})= 2\varepsilon (b_{1g}^{a})+ U + \Delta U,
$$
with $U$ being the contribution of the  hole-hole $b_{1g}^{b}-b_{1g}^{b}$ interaction, and $\Delta U$  that of configurational interaction, then
$$
             U + \Delta U = 4.7 -3.5 = 1.2 (eV),
$$
which value is puzzlingly small as compared with  the bare value $U_{d}= A +4B +3C \simeq 10$ eV for purely atomic  $d_{x^{2}-y^{2}}^{2}$ configuration.

A concept of the well isolated Zhang-Rice singlet to be a ground state of the
two-hole $CuO_{4}^{5-}$ center, is a guideline of many popular model approaches
\cite{Dagotto}. Namely with ZR singlet one associates the first
electron-removal state and the lowest-lying features in ARPES spectra. At the
same time, a number of experimental data and theoretical models evidence a more
complicated structure of the valent multiplet for the two-hole  $CuO_{4}^{5-}$
center.

A model of the valent $^{1}A_{1g}-{}^{1}E_{u}$ multiplet, developed in Refs.
\cite{Moskvin94,Moskvin97,Moskvin98}, implies a quasi-degeneracy in the ground
state of the two-hole $CuO_{4}^{5-}$ center with two close in energy
$^{1}A_{1g}$ and ${}^{1}E_{u}$ terms of $b_{1g}^2$ and $b_{1g}e_u$
configurations, respectively. In other words, one implies two near equivalent
allocations for the additional hole, either to the $Cu3dO2p$ hybrid   $b_{1g}$
state, or to purely oxygen $e_u$ state with peculiar $Cu^{2+}-Cu^{3+}$ valence
resonance (see Fig.1).   Occurrence of the localized purely oxygen $e_u$ like
states is provided, in particular, by the specific properties of the
''non-rigid'' anionic $O2p^6$ background \cite{Hirsch,background}.

The model approach under consideration is based both on microscopic quantum
chemical study of the model copper-oxygen clusters
\cite{Moskvin94,Tanaka,Tanaka1} and a large variety of experimental data. To
the best of our knowledge one of the first quantitative conclusions on the
competitive role of the hybrid copper-oxygen $b_{1g}(d_{x^2 -y^2})$ orbital and
purely oxygen $O2p_{\pi}$ orbitals in formation of valent states near the Fermi
level in
 the $CuO_2$ planes has been made by Jiro Tanaka {\it et al.} \cite{Tanaka}
 (see also more later publication \cite{Tanaka1}).

 In a sense, the valent $(b_{1g}^{2}){}^{1}A_{1g}-(b_{1g}e_{u}){}^{1}E_{u}$ manifold
for the hole $CuO_{4}^{5-}$ center implies an unconventional state with $Cu$
valence resonating
 between $Cu^{3+}$ and $Cu^{2+}$, or "ionic-covalent" bonding \cite{Goodenough}.
 In fact, the $CuO_4$ center with the valent
 $(b_{1g}^{2}){}^{1}A_{1g}-(b_{1g}e_{u}){}^{1}E_{u}$ manifold represents
 a specific version of the "correlation" polaron, introduced by
 Goodenough and Zhou \cite{Goodenough}.

 The model allows to consistently explain many puzzling properties both of
insulating and superconducting cuprates: the mid-infrared (MIR) region
absorption bands \cite{Moskvin94}, the (pseudo)Jahn-Teller effect and related
phenomena \cite{Moskvin97}, the spin properties \cite{Moskvin98}.

The presence of small polarons in semiconducting copper oxides has been
detected with photoinduced infrared absorption measurements, infrared
spectroscopy, and X-ray absorption fine structure techniques.

One of the most exciting experimental evidences in favour of the model with the
valent $^{1}A_{1g}-{}^{1}E_{u}$ multiplet is associated with observation in the
doped cuprates of the MIR bands which polarisation features are compatible with
those for $^{1}A_{1g}-{}^{1}E_{u}$ intra-multiplet dipole transitions
\cite{Moskvin94}. The corresponding transition energies observed for various
cuprates are of the order of a few tenths of eV, that exhibits a typical energy
scale for the valent multiplet.

The $e_u$ hole can be coupled with the $b_{1g}$ hole both antiferro- and
ferromagnetically. This a rather simple consideration indicates clearly a
neccessity to incorporate in valent multiplet both spin-singlet
$(b_{1g}e_{u}){}^{1}E_{u}$ and spin-triplet $(b_{1g}e_{u}){}^{3}E_{u}$, which
energy could be even lower due to ferromagnetic   $b_{1g}-e_{u}$ exchange.
Indeed, the low-lying spin-triplet state for the two-hole  $CuO_{4}^{5-}$
center was revealed with the help of $^{63,65}Cu$ NQR in
$La_{2}Cu_{0.5}Li_{0.5}O_{4}$ with singlet-triplet separation  $\Delta_{ST} =
0.13$ eV \cite{Fisk}. Indirect manifestation of  $O2p\pi$, or $e_u$ valent
states  were detected in Knight shift measurements by NMR for 123-YBaCuO system
\cite{Yoshinari}. In connection with the  valent $^{1}A_{1g}-{}^{1}E_{u}$
multiplet model for copper oxides one should note and comment the results of
paper by Tjeng et al. \cite{Tjeng}, where the authors state that they ''are
able to unravel the different spin states in the single-particle excitation
spectrum of antiferromagnetic  $CuO$ and show that the top of the valence band
is of pure singlet character, which provides strong support for the existence
and stability of Zhang-Rice singlets in high-$T_c$ cuprates''. However, in
their photoemission work they made use of the $Cu2p_{3/2}(L_{3})$ resonance
condition that allows to detect unambiguously only copper photohole states,
hence they cannot see the purely oxygen photohole $e_u$ states.

It should be noted that the complicated $^{1}A_{1g}-{}^{1,3}E_{u}$ structure of
the valent multiplet for the two-hole $CuO_{4}^{5-}$ center has to be revealed
in the photoemission spectra, all the more that the odd ${}^{1,3}E_{u}$ terms
play a principal role: namely these yield a nonzero contribution to ARPES for
${\bf k}=0$, or, in other words, at  $\Gamma$ point. In this connection one
should note experimental measurements of the photoemission spectra in
$Sr_2CuO_2Cl_2$ \cite{Wells} and $Ca_2CuO_2Cl_2$ \cite{Ronning}. All these
clearly detect a nonzero photocurrent intensity in the BZ center, thus
supporting the $^{1}A_{1g}-{}^{1,3}E_{u}$ structure of the ground state valent
multiplet.

\vspace{0.4cm}

{\bf 1.2. Effective vibronic Hamiltonian and small (pseudo)Jahn-Teller polaron}

\vspace{0.2cm}

The orbital near-degeneracy leads to strong electron-lattice (vibronic) effects.
For  $(^{1}A_{1g},^{1}E_{u})$ valent multiplet in the $CuO_4$-center  one has to consider the active  $Q_{\gamma}$ nuclear displacement modes with  $\gamma = a_{1g}$, $b_{1g} $, $b_{2g}$, $e_{u}$.
Effective vibronic Hamiltonian $\hat H_{vibr}$ is a sum of two contributions
\begin{equation}
\hat H_{vibr}=\hat H_{EE}+\hat H_{AE},
\end{equation}
that of vibronic coupling within the $^{1}E_{u}$ term, and as well the vibronic
coupling of the  $^{1}E_{u}$ and $^{1}A_{1g}$ terms, respectively. Vibronic
Hamiltonian for isolated $^{1}E_{u}$ term is familiar for a so called
$E-b_{1}-b_{2}$ problem \cite{ber}, and in terms of orbital operators ${\hat
V}_{\gamma}$ ($\gamma =a_{1g},b_{1g},b_{2g},e_u$) has a following form:
\begin{equation}
\hat H_{EE} =v_{b_{1g}}{\hat V}_{b_{1g}}Q_{b_{1g}} +v_{b_{2g}}{\hat V}_{b_{2g}}Q_{b_{2g}}
\end{equation}
with $v_{b_{1g},b_{2g}}$ being vibronic coupling parameters.
Vibronic Hamiltonian $\hat H_{AE}$ contains two terms
\begin{equation}
\hat H_{AE}=v_{a_{1g}}{\hat V}_{a_{1g}}Q_{a_{1g}}+\sum_{e_{u}}v_{e}({\vec
Q}_{e_{u}}^{i}{\hat {\vec V}_{e_{u}}}).
\end{equation}
The first provides the connection between the  $A-E$ separation and symmetrical
$Q_{a_{1g}}$ mode, while the second describes a linear vibronic coupling of the
  $^{1}E_{u}$ and $^{1}A_{1g}$ terms due to the active odd
  $Q_{e_{u}^{x}},Q_{e_{u}^{y}}$ modes. One should note that for the
  $CuO_4$ center one exists three types of the  $e_{u}$ modes.
  Vibronic Hamiltonian $\hat{H}_{vib}$ has to be added by elastic energy $\hat{U}_{Q}$ for the $CuO_4$ center
\begin{equation}
\hat{U}(Q)=\sum_{i}\frac{\omega _{i}^{2}Q_{i}^{2}}{2},
\end{equation}
where summing runs over all normal displacements modes. A rather complete
examination of the (pseudo)Jahn-Teller
$(^{1}A_{1g},^{1}E_{u})$-$a_{1g}$-$b_{1g}$-$b_{2g}$-$e_{u}$ problem was carried
out in Ref. \cite{Moskvin97}. Depending on the relation between vibronic and
elastic parameters one finds to be three situations:

1. Weak vibronic coupling when $\hat{H}_{vib}$  results only in a rather small renormalization of bare elastic constants.

2.  The Jahn-Teller effect for the  $E_u$ term ($E-b_1-b_2$ problem) with emergence of two-well adiabatic potential for the rhombic modes with either $b_{1g}$, or $b_{2g}$ symmetry.

3.  Strong pseudo-Jahn-Teller effect with vibronic mixing of $(^{1}A_{1g}$ and $^{1}E_{u})$ terms and emergence of four-well adiabatic potential.

In practice one might observe various manifestations of the (pseudo)Jahn-Teller effect including the conventional static or dynamic effect, local structural instability, including the dipole one, spontaneous and induced local structural phase transitions with reconstruction of the multi-well adiabatic potential.

The presence of the PJT centers, or small PJT polarons in semiconducting copper
oxides has been detected, in particular, with photoinduced infrared absorption
measurements, infrared spectroscopy, and X-ray absorption fine structure
techniques. The photoinduced infrared absorption measurements on a number of
insulating copper oxides have unambiguously revealed the formation of a
localized electronic state accompanied by a localized structural distortion.
These two aspects of the data have demonstrated the self-localized polaronic
nature of the photo-injected carriers with a  complex optical response
spreading on a rather wide spectral range $0.1\div 1.0$ eV.


\vspace{0.4cm}
{\bf 1.3. Large non-adiabatic hole polarons in the $CuO_2$ layers of cuprates}
\vspace{0.2cm}

Non-zero hole transfer together with strong intercenter coupling due to the common oxygen favours the transformation of the small PJT polaron into a large non-adiabatic PJT polaron. Stabilization of a large polaron causes the hole to be shared over several $CuO_4$ centers that could result in an optimal relaxation of the elastic energy and vibronic coupling all over the polaron volume.

A model of large non-adiabatic hole polarons for the doped copper oxides was proposed by Bersuker and Goodenough \cite{GB}. The large polaron could contain a hole cloud distributed over a significant number of the $CuO_4$ centers. So, the authors \cite{GB} considered polarons containing 5 to 6 $CuO_4$ centers.

 One of the model approaches capable to effectively describe such a polaron implies an extention of the conventional PJT problem with  choice of a set of the symmetrized quasimolecular orbitals constructed from the valent hole states localized on the $CuO_4$ clusters involved in the PJT polaron (similar to the known LCAO-method), and a set of  symmetrized displacements for the corresponding copper and oxygen atoms.

The symmetrized quasimolecular orbitals can be build as linear combinations of
the valent hole molecular orbitals $\psi_{\gamma\mu}({\bf r})$  for the $CuO_4$
cluster
\begin{equation}
\Psi^{(\gamma\mu)}_{\Gamma M}({\bf R}) = \sum_{{\bf r}} C^{\Gamma
M}_{\gamma\mu} ({\bf R}+{\bf r}) \psi_{\gamma\mu}({\bf R}+{\bf r})\psi_{0}({\bf
R}+{\bf r}),
\end{equation}
where $\psi_{0}({\bf r})$ is ground state wave function of the $CuO^{6-}_{4}$
cluster, $\Gamma M, \gamma \mu$ are irreducible representations of the point
symmetry group, $C^{\Gamma M}_{\gamma\mu}$ are the symmetry coefficients, ${\bf
R}$ is a radius-vector of the polaronic center of symmetry, and the sum runs
over all the $CuO_4$ centers inside large polaron. For illustration we present
the symmetrized quasimolecular orbitals for the large polaron containing 5
$CuO_4$ centers (see Fig.2) with valent $b_{1g}$ orbital:
$$
\Psi^{(b_{1g})}_{A_{1g}}=\chi({\bf 0}),
\Psi^{(b_{1g})}_{A_{1g}^{'}}=\frac{1}{2}\{ \chi({\bf x})+\chi({\bf
-x})+\chi({\bf y})+\chi({\bf -y}) \},
$$
$$
\Psi^{(b_{1g})}_{B_{1g}}=\frac{1}{2}\{ \chi({\bf x})+\chi({\bf -x})-\chi({\bf
y})-\chi({\bf -y}) \},
$$
$$
\Psi^{(b_{1g})}_{E_{u1}}=\frac{1}{2}\{ \chi({\bf x})-\chi({\bf -x})+\chi({\bf
y})-\chi({\bf -y}) \},
$$
$$
\Psi^{(b_{1g})}_{E_{u2}}=\frac{1}{2}\{-\chi({\bf x})+\chi({\bf -x})+\chi({\bf
y})-\chi({\bf -y}) \},
$$
where $|\chi\rangle = |\psi_{b_{1g}}\psi_{b_{1g}}\rangle \equiv
|\psi_{0}\psi_{0}\rangle$ is the Zhang-Rice singlet wave function.

 Thus, we obtain a polaronic $A_{1}^{1}-A_{1}^{2}-B_{1}-E$ manifold generated by the $b_{1g}$-state of the $CuO_4$ centers. This manifold incorporates  two quasimolecular orbitals with $s$-symmetry,
one with $d_{x2-y2}$-symmetry, and doublet with $p_{x,y}$-symmetry. The account for the hole transfer results in a bare splitting within the manifold.
The orbital states with the same symmetry will mix forming the superpositions
\begin{equation}
\Phi_{\Gamma M}({\bf R}) = \sum_{\gamma\mu}\sum_{{\bf
r}}A^{(\gamma\mu)}_{\Gamma M} C^{\Gamma M}_{\gamma\mu} ({\bf R}+{\bf r})
\psi_{\gamma\mu}({\bf R}+{\bf r})\psi_{0}({\bf R}+{\bf r}). \label{polaron_wf}
\end{equation}

One should note a specific bare quasimolecular orbitals for the $N$-center
''molecule'' generated by local current states like  $e_{u\pm}\propto
(e_{ux}\pm ie_{uy})$. These manifest the $N$-fold pattern of microscopic
circulating currents resulting both in a state with nonzero magnetic moment and
in a state with zero magnetic moment. In the latter case the time-reversal
symmetry as well as rotational symmetry is broken but the product of the two is
conserved.

For many applications in  scattering problems it is useful to introduce the
form-factors
\begin{equation}
f^{(\gamma\mu)}_{\Gamma M}({\bf k}) = \sum_{{\bf r}}A^{(\gamma\mu)}_{\Gamma M}
C^{\Gamma M}_{\gamma\mu}({\bf r})e^{i{\bf k}{\bf r}} \label{formfactor}
\end{equation}
to describe the spatial distribution of the hole density inside a large
polaron.

Some form-factors for $N=4$ large polaron (see Fig.2) are

$$
f^{(b_{1g})}_{B_{1g}}({\bf k})=2\cos(\frac{ak_{x}}{2})\cos(\frac{ak_{y}}{2}),
$$

$$
f^{(e_{u}x)}_{B_{1g}}({\bf
k})=-i\sqrt{2}\sin(\frac{ak_{x}}{2})\cos(\frac{ak_{y}}{2}), \quad
f^{(e_{u}y)}_{B_{1g}}({\bf
k})=i\sqrt{2}\cos(\frac{ak_{x}}{2})\sin(\frac{ak_{y}}{2}),
$$

$$
f^{(b_{1g})}_{E_{u}x}({\bf k})= 2 i
\sin(\frac{ak_{x}}{2})\cos(\frac{ak_{y}}{2}), \quad f^{(b_{1g})}_{E_{u}y}({\bf
k})=2i\sqrt{2}\cos(\frac{ak_{x}}{2})\sin(\frac{ak_{y}}{2}),
$$

$$
f^{e_{u}x}_{E_{u}x}({\bf k})=f^{e_{u}y}_{E_{u}y}({\bf
k})=2\cos(\frac{ak_{x}}{2})\cos(\frac{ak_{y}}{2}), \quad
f^{e_{u}y}_{E_{u}x}({\bf k})=f^{e_{u}y}_{E_{u}x}({\bf k})=0.
$$

A set of the symmetrized quasimolecular orbitals $\Phi_{\Gamma M}$ and
symmetrized displacements $Q_{\Gamma M}$ for the  copper and oxygen atoms
inside large polaron form a basis for a rather standard, albeit complicated,
vibronic PJT problem which solution gives an energy spectrum and appropriate
wave functions $\Psi_{\alpha \Gamma n}$ describing strongly correlated vibronic
nature of the polaronic states. Here, $\alpha \Gamma n$ represent a set of
quantum numbers which define a vibronic state. It should be noted, that only
given the extremely simplifying assumptions this function could be written in a
familiar Born-Oppenheimer form like $\Psi_{\alpha \Gamma n}=\Phi_{\Gamma M}\chi
_{\alpha \Gamma n}(Q)$, where $\chi _{\alpha \Gamma n}(Q)$ is a vibrational
function. In a wide sense, excitation spectrum of the large PJT polaron
involves a whole vibronic spectrum originated from quasimolecular
$(\gamma)\Gamma$ multiplet. It should be noted that the symmetry classification
of polaronic hole states allows to elucidate some similarities with atomic
system and partial waves; one might  say about $s-,p-,d-,...$ like hole states.
One should note an occurrence of the time-reversal symmetry breaking current
states for the large PJT polaron. In addition, we must emphasize one more  the
specific role of the near-degeneracy for the valent manifold, and probable PJT
effect, for the individual $CuO_4$ center in formation of a large non-adiabatic
polaron.

Finally, large PJT-polaron in the lattice could be represented as a system of
the $CuO_4$ centers with a set of metastable states $\Psi_{\alpha \Gamma n}$
specified by a binding energy and a life-time. Its nature implies strong charge
fluctuations, so, it seems rather difficult to confine such a polaron within a
single $CuO_2$ layer. In other words, the large PJT polaronic nature of
photo-hole implies its $3D$ structure with finite dimension, or correlation
length,  in the $c$-direction. Three-dimensional structure of the large PJT
polaron could result in a rather strong $k_{\perp}$, and consequently,
photon-energy dependence of the ARPES intensity. Nevertheless, below we
restrict ourselves, for simplicity, with the planar PJT polarons.

 In practice, for real systems the polaron will effectively
couple with all the phonon modes which are active in the PJT effect \cite{ber}.
In other words, a large non-adiabatic PJT polaron may be considered
as  a bounded state of the large PJT-center and phonons. Coupling with phonons and
spin system will result in an effective enlargement of polaron. The simplest way to
account for this effect implies the introducing of a momentum-dependent cut-off factor
 to a polaronic form-factor. In addition, the phonon system will determine the relaxation
 dynamics of polaron states, and gives rise an effective dispersion to
the hole spectral function.

It should be noted that the quasiparticle behavior of polaron implies, as a
rule, a single rather strongly bounded long-lived term. For this one can
introduce an effective Hamiltonian which should include polaronic transport and
coupling to phonon and spin lattice modes thus providing the coherent and
incoherent part of the quasiparticle spectral function. It should be noted that
the effective quasiparticle Hamiltonian could be look like as familiar Hubbard,
or $t-J$ Hamiltonian.  The remaining short-lived polaronic states will give
rise to a rather wide structureless  and dispersionless background in spectral
region of valent manifold.

Above we have considered the large lattice polaron and ignored the role of the
antiferromagnetic background and appropriate spin fluctuations. Many authors
have considered the formation of magnetic (spin) polarons \cite{Dionne},
moreover the assumption of spin-polaronic nature of the photo-hole, described
in the framework of the extended $t-J$ model is one of the most popular
approaches to the interpretation of ARPES data in cuprates. So-called
electron-hole asymmetric small polarons were introduced by J.E. Hirsch
\cite{Hirsch}. In general, the polarons must be of complex spin-lattice hybrid
type with a complicated spatial distribution of the electron and  spin
densities, and local structure distortions.

\newpage

\newpage
\vspace{0.4cm} {\large \bf 2. Expression for photointensity} \vspace{0.2cm}

 Below we consider an expression for intensity of the photoemission
with creation of immobile hole PJT polaron. An effective Hamiltonian for
interaction  with the electromagnetic field of frequency $\omega $ and
polarisation ${\bf e}$ could be written within a polaronic manifold as follows
\begin{equation}
\hat{H}_{int} = \sum_{\Gamma M}\sum_{{\bf k}} {\cal{M}}_{\Gamma M}({\bf k},{\bf
e}) \hat{c}^{\dag}_{{\bf k}\sigma} \hat{h}^{\dag}_{\Gamma M \sigma} + H.c.,
\end{equation}
where ${\bf k}$ is a momentum of the final state of the photoelectron
registered by the detector, $\hat{c}^{\dag}_{{\bf k}\sigma}$ and
$\hat{h}^{\dag}_{\Gamma M\sigma}$ are
 creation operators for photoelectron and photohole, respectively.
The matrix element is given by
\begin{equation}
{\cal{M}}_{\Gamma M}({\bf k},{\bf e}) =\langle \psi _{{\bf k}}({\bf r})
\Psi^{(N-1)}_{\Gamma M} |\hat{H}_{eR}| \Psi^{(N)}_{g}\rangle, \label{1}
\end{equation}
where
$$
\hat{H}_{eR}=\frac{e\hbar}{2mc}({\bf{p}}\cdot {\bf{A}}+{\bf{A}}\cdot {\bf{p}})
$$
is the interaction Hamiltonian with the electron momentum operator ${\bf{p}}$
and the vector potential ${\bf{A}}$ of the photon field; $\Psi^{(N)}_{g}$ is
the wave function for the ground state; $\Psi^{(N-1)}_{\Gamma M }$ is the wave
function for a $\Gamma M$ state with one removed electron (one additional
hole); $\psi _{{\bf k}}({\bf r})$ is the photoelectron wave function. It should
be noted that expression (\ref{1}) already implies a number of noticeable
simplifications.

Modelling the photoelectron wave function by a plane wave, we rewrite the
expression for the matrix element (\ref{1})  as follows
\begin{equation}
{\cal{M}}_{\Gamma M}({\bf k},{\bf e})=\sum_{\gamma\mu}f^{(\gamma\mu)}_{\Gamma
M}({\bf k}) M_{\gamma\mu}({\bf k},{\bf e}), \label{4}
\end{equation}
where in the dipole approximation
\begin{equation}
M_{\gamma\mu}({\bf k},{\bf e})=\langle \psi_{\gamma\mu}({\bf r})|( {\bf{e}}
\cdot {\bf r} )|e^{i{\bf k}{\bf r}}\rangle. \label{7a}
\end{equation}

Finally, the expression for the photoemission intensity may be transformed into
\begin{equation}
I({\bf k},\omega,{\bf e})\propto
\\
\sum_{\Gamma _{1} M_{1};\Gamma _{2} M_{2}} {\cal{M}}^{*}_{\Gamma _{1}
M_{1}}({\bf k},{\bf e}) {\cal{M}}_{\Gamma _{2} M_{2}}({\bf k},{\bf e})
A_{\Gamma _{1} M_{1};\Gamma _{2} M_{2}}(\omega), \label{4}
\end{equation}
where the ground $|g\rangle$ and excited $|e\rangle$   states are the
nonperturbed electron-vibrational states for the N-center cluster and vibronic
ones for the hole PJT polaron, respectively. Emission spectral functions have a
quite standard form
$$
A_{\Gamma _{1} M_{1};\Gamma _{2} M_{2}}(\omega)=\frac{1}{2}\sum _{\sigma ,
e,g}e^{-\beta E_{g}}\langle e|\hat{h}^{\dag}_{\Gamma _{1} M_{1}\sigma}|g\rangle
\langle g|\hat{h}_{\Gamma _{2} M_{2}\sigma}|e\rangle \delta (\omega
+E_{e}-E_{g})=
$$
\begin{equation}
\frac{1}{2}\sum _{\sigma}\int dt e^{i\omega t}\langle \hat{h}^{\dag}_{\Gamma
_{1} M_{1}\sigma}(t)\hat{h}_{\Gamma _{2} M_{2}\sigma}(0)\rangle.
\end{equation}
Spectral functions contain a complete information about complex vibronic
structure of the PJT polaron, and  describe both the partial
$\Gamma$-contributions at $\Gamma _{1}=\Gamma _{2}$ and interference effects
for different states with the same symmetry. These obey the sum rules
\begin{equation}
\int \frac{d \omega}{2\pi}A_{\Gamma _{1} M_{1};\Gamma _{2} M_{2}}(\omega)=
n_{\Gamma _{1} M_{1}} \delta _{\Gamma _{1} M_{1};\Gamma _{2} M_{2}}.
\end{equation}

Despite the extremely rough simplifying approximations Exp.(\ref{4}) displays
very complex multi-level structure for photo-intensity with nontrivial
polarization and  ${\bf k},\omega$ dependence.

Calculation of spectral functions $A_{\Gamma _{1} M_{1};\Gamma _{2}
M_{2}}(\omega)$ for the large PJT polaron represents an extremely complex
problem even at very strong simplifications \cite{ber}. For illustration, one
might refer to a similar problem with spectral function which describes the
line-shape of the optical $A-E$ transition between the orbital singlet and
orbital JT doublet \cite{ber}. 

It should be noted, that, in a sense, the non-diagonal spectral functions
$A_{\Gamma _{1} M_{1};\Gamma _{2} M_{2}}(\omega)$ describe the spectral weight
transfer between $\Gamma _{1}$ and $\Gamma _{2}$ bands. In a whole, the PJT
polaronic nature of the photo-hole provides a ${\bf k}$-dependent ARPES
spectral line-shape, in particular, with dispersive peak position.

\vspace{0.4cm} {\large \bf 3. One-electron matrix element} \vspace{0.2cm}

\vspace{0.4cm} {\bf 3.1. Copper contribution} \vspace{0.2cm}

The copper atomic orbital with symmetry $\gamma\mu$ can be represented in the
form

$$
d_{\gamma\mu}({\bf r})=R_{3d}(r)\sum_{m} \alpha_{2m}(\gamma\mu) Y_{2m}({\bf
r}),
$$
where $\alpha_{2m}(\gamma\mu)$ are coefficients specified by the symmetry
requirements, $R_{3d}(r)$ radial wave function, which we assume to be of simple
Slater form
$$
R_{3d}(r)=\frac{2}{81}\sqrt{\frac{2}{15}}\frac{r^{2}}{a^{3}_{d} \sqrt{a_{d}}}
exp\{ -\frac{r}{3a_{d}} \}.
$$
Inserting these expressions to (\ref{7a}) and making use the familiar expansion
for the plane wave \cite{Varshalovich}

\begin{equation}
e^{i{\bf
k}{\bf{r}}}=4\pi\sum_{L=0}^{\infty}\sum_{M=-L}^{L}i^{L}j_{L}(kr)Y^{*}_{LM}({\bf
k}) Y_{LM}({\bf{r}}) \label{exp}
\end{equation}
we rewrite (\ref{7a}) as follows

$$
M^{(Cu)}_{\gamma\mu}({\bf k}, {\bf{e}}) = \langle
d_{\gamma\mu}({\bf{r}})|({\bf{e}} \cdot {\bf{r}}) |e^{i{\bf k}{\bf{r}}}
\rangle=
$$
\begin{equation}
 \frac{4\pi i}{\sqrt{5}} \{
\sqrt{2}D_{1}(k)K^{(\gamma\mu)}_{1}({\bf{e}},{\bf k})+
\sqrt{3}D_{3}(k)K^{(\gamma\mu)}_{3}({\bf{e}},{\bf k})\},
\end{equation}
where we denote

$$
D_{1}(k)= \langle R_{3d}(r)|r| j_{1}(kr)  \rangle = 864 \sqrt{\frac{6}{5}}
\frac{ a^{3}_{d} \sqrt{a_{d}} k (5-27
a^{2}_{d}k^{2})}{(1+9a^{2}_{d}k^{2})^{5}},
$$

$$
D_{3}(k)= \langle R_{3d}(r)|r| j_{3}(kr)  \rangle = 62208 \sqrt{\frac{6}{5}}
\frac{ a^{5}_{d} \sqrt{a_{d}} k^{3}}{(1+9a^{2}_{d}k^{2})^{5}},
$$

$$
K^{(\gamma\mu)}_{L}({\bf{e}},{\bf k})=\bigl[ Y^{L}\times e^{1} \bigr] ^{2\gamma
\mu *}= \sum_{M,q,m}(-1)^{q}e_{-q}C^{2m}_{LM1q} Y^{*}_{LM}({\bf
k})\alpha^{*}_{2m}(\gamma\mu).
$$
Here $C^{2m}_{LM1-q}$ are the Clebsch-Gordan coefficients. The photoelectron
energy dependence of quantities $D_{1}(k)$ and $D_{3}(k)$ given the  $Cu3d$
radial parameter $a_{d}=0.35$ \AA\quad is shown in Fig.3.

\vspace{0.4cm} {\bf 3.2. Oxygen contribution} \vspace{0.2cm}

The oxygen molecular orbital can be represented as a linear combination of
atomic  $O2p$ functions centered at appropriate oxygen positions
\begin{equation}
p_{\gamma \mu}({\bf {r}})=\sum_{{\bf{t}} m} C_{m}^{\gamma \mu}({\bf{t}})
R_{2p}(|{\bf {r}}-{\bf{t}}|)Y_{1m}({\bf{r}}-{\bf{t}}), \label{9}
\end{equation}
where $C_{m}^{\gamma \mu}({\bf{t}})$ are coefficients specified by the symmetry
requirements, $R_{2p}(r)$ radial wave function. Inserting (\ref{9}) to
(\ref{7a}) and making substitution ${\bf{r'}}={\bf{r}}-{\bf{t}}$, we reduce
(\ref{7a}) to

\begin{eqnarray}
M({\bf k},{\bf{e}}) &=& \sum_{{\bf{t}} m} C^{*}_{m}({\bf{t}})({\bf {e}}\cdot
{\bf{t}})e^{i{\bf
 k}{\bf{t}}}\langle
R_{2p}({\bf{r'}})Y_{1m}({\bf{r'}})|e^{i{\bf k}{\bf{r'}}}\rangle \cr &+&
\sum_{{\bf{t}} m} C^{*}_{m}({\bf{t}})e^{i{\bf
 k}{\bf{t}}}\langle R_{2p}(r')Y_{1m}({\bf{r'}})|({\bf
{e}}\cdot{\bf{r'}})|e^{i{\bf k}{\bf{r'}}}\rangle. \label{10}
\end{eqnarray}
For convenience, one introduces two vectors with cyclic components
\begin{equation}
G_{m}({\bf k},{\bf{e}})=\sum_{{\bf{t}}} C^{*}_{m}({\bf{t}})({\bf {e}}\cdot
{\bf{t}})e^{i{\bf  k}{\bf{t}}}, \quad Z_{m}({\bf k})=\sum_{{\bf{t}}}
C^{*}_{m}({\bf{t}})e^{i{\bf
 k}{\bf{t}}}. \label{11}
\end{equation}
Then Exp.(\ref{10}) could be rewritten in a more compact form

$$ M({\bf k},{\bf{e}}) = \sum_{m}G_{m}({\bf k},{\bf{e}})\langle
R_{2p}(r)Y_{1m}({\bf{r}})|e^{i{\bf k}{\bf{r}}}\rangle + $$
\begin{equation}
\sum_{m}Z_{m}({\bf k})\langle R_{2p}(r)Y_{1m}({\bf{r}})|({\bf
{e}}\cdot{\bf{r}})|e^{i{\bf k}{\bf{r}}}\rangle. \label{12}
\end{equation}
Making use the familiar expansion (\ref{exp}) for the plane wave we rewrite
(\ref{12}) as follows

$$
M^{(O)}_{\gamma\mu}({\bf k},{\bf{e}})=2\sqrt{3\pi} i
B(k)({{\bf{G}}}_{\gamma\mu}\cdot {\bf k})/k   +
$$
\begin{equation}
 \sqrt{\frac{4\pi}{3}} \{
A_{0}(k)({{\bf{Z}}}_{\gamma\mu}\cdot{\bf{e}})- A_{2}(k) \frac{ 3({\bf{e}} \cdot
{\bf k}) ({{\bf{Z}}}_{\gamma\mu}\cdot{\bf k})-k^{2}({{\bf{Z}}}_{\gamma\mu}
\cdot {\bf{e}}) }{k^{2}} \} \label{13}
\end{equation}
where we add the $\gamma\mu$ indexes label for the molecular orbital under
consideration. Rather simple analytical expressions for the radial integrals
$A_{0,2},B$ in (\ref{13}) could be obtained if to make use the simplest Slater
$O2p$ radial wave function
\begin{equation}
R_{2p}(r)=\frac{1}{2{\sqrt{6}}}\frac{1}{{\sqrt{{a_{
p}^{3}}}}}\frac{r}{a_{p}}{exp\{-\frac{r}{2a_{p}}\}}. \label{14}
\end{equation}
Then
$$ B(k)=\langle R_{2p}(r)| j_{1}(kr)  \rangle=\frac{64 \sqrt{6}
{a_{p}^{2}\sqrt{a_{p}}} k}{{ 3 {(1+4 {a_{p}^2} {k^2})}^3}},
$$
$$
A_{0}(k)=\langle R_{2p}(r)|r| j_{0}(kr)  \rangle=\frac{64 \sqrt{6} a_{p}^{2}
{\sqrt{a_{p}} }(1 - 4 {a_{p}^2} {k^2})}{{{(1 + 4 {a_{p}^2} {k^2})}^4}},
$$
$$ A_{2}(k)=\langle R_{2p}(r)|r| j_{2}(k\rho)
\rangle=\frac{512 \sqrt{6} a_{p}^{4} \sqrt{a_{p}} {k^2}}{{{(1+4 {a_{p}^2}
{k^2})}^4}}.
$$
The photoelectron energy dependence of quantities $B(k)$, $A_{0}(k)$ and
$A_{2}(k)$ given the  $O2p$ radial parameter $a_{p}=0.52$ \AA\quad is shown in
Fig.4.

\vspace{0.4cm}

{\bf 3.3. Expression for $M_{b_{1g}}({\bf k},{\bf{e}})$}

\vspace{0.2cm}

Below we consider in detail the matrix element which specifies the contribution
of the electron-removal process from the $b_{1g}$ orbital to form the
Zhang-Rice singlet.

For $\gamma\mu = b_{1g}$ at rather large photon energy $E_{ph}>20$ $eV$, but
small binding energy $E<1\div2$ $eV$)

\begin{equation}
M^{(Cu)}_{b_{1g}}({\bf k}, {\bf{e}})=2 i\sqrt{\frac{3\pi}{5}}\{
D_{1}(k)+\frac{7}{2}D_{3}(k) \} (e_{x}\kappa _{x} - e_{y}\kappa _{y}),
\end{equation}
where $\kappa_{\alpha} = k_{\alpha}/k$, $\alpha=x,y,z$. Making use the
numerical values for the coefficients $C^{(b_{1g})}_{m}({\bf{t}})$ from Table 1
one might obtain

\begin{eqnarray}
M^{(O)}_{b_{1g}}({\bf k}, {\bf{e}}) =
 i\sqrt{\frac{\pi}{3}} \{
-3aB(k)[\kappa_{x}e_{x}\cos(\frac{ak_{x}}{2})-
\kappa_{y}e_{y}\cos(\frac{ak_{y}}{2})] -  \nonumber \\
2[A_{0}(k)+A_{2}(k)][e_{x}\sin(\frac{ak_{x}}{2})-e_{y}\sin(\frac{ak_{y}}{2})] +
\nonumber \\
6A_{2}(k)[e_{x}\kappa_{x}+e_{y}\kappa_{y}][\kappa_{x}\sin(\frac{ak_{x}} {2}) -
\kappa_{y}\sin(\frac{ak_{y}}{2})] \}. \label{15}
\end{eqnarray}
 It should be emphasized
that the photocurrent intensity at the BZ center in the case of the even
$b_{1g}$ orbital as well as for any other even $\gamma$ orbital turns to zero.

 The bonding one-electron molecular $b_{1g}$ orbital can
be written as follows
$$
\Psi_{b_{1g}}({\bf{r}})=d_{b_{1g}}({\bf{r}})\sin\theta_{b_{1g}} +
p_{b_{1g}}({\bf{r}})\cos\theta_{b_{1g}},
$$
where  $\theta_{b_{1g}}$ is an angular covalent mixing parameter. Then the
photocurrent intensity will be proportional to
$$
|\langle\Psi_{b_{1g}}({\bf k})|({\bf{e}}\cdot {\bf{r}})| e^{i{\bf
k}{\bf{r}}}\rangle|^{2} = |M^{(Cu)}_{b_{1g}}|^{2}\sin^{2}\theta_{b_{1g}} +
$$
\begin{equation}
|M^{(O)}_{b_{1g}} |^{2} \cos^{2}\theta_{b_{1g}} + \sin \theta_{b_{1g}}
\cos\theta_{b_{1g}} \{
(M^{(Cu)}_{b_{1g}})^{*}M^{(O)}_{b_{1g}}+M^{(Cu)}_{b_{1g}}(M^{(O)}_{b_{1g}})^{*}
\} .
\end{equation}

\vspace{0.4cm} {\bf 3.4. Expression for $M_{e_{u}}({\bf k}, {\bf{e}})$}
\vspace{0.2cm}

As it was mentioned above, the oxygen  $e_{u}$ states one might subdivide to
$\sigma$ and  $\pi$ orbitals. Due to  a strong $O2p-O2p$ coupling they
hybridize to form bonding and antibonding molecular $e_{u}$ orbitals
\cite{Moskvin94}. The bonding one-electron molecular $e_{u}$ orbital can be
written as follows
$$
\Psi_{e_{u}\mu}({\bf{r}})=p^{(\pi)}_{e_{u}\mu}({\bf{r}})\sin\theta_{e_{u}} +
p^{(\sigma)}_{e_{u}\mu}({\bf{r}})\cos\theta_{e_{u}}, \qquad \mu=x,y,
$$
where  $\theta_{e_{u}}$ is a covalent mixing parameter. Then the photocurrent
intensity will be proportional to
$$
\sum_{\mu}|\langle\Psi_{e_{u}\mu}({\bf{r}})|( {\bf{e}} \cdot {\bf{r}} )|
e^{i{\bf k}{\bf{r}}}\rangle|^{2} =
\sum_{\mu}|M^{(\pi)}_{\mu}|^{2}\sin^{2}\theta_{e_{u}} +
$$
\begin{equation}
 \sum_{\mu}|M^{(\sigma)}_{\mu}|^{2} \cos^{2}\theta_{e_{u}}+ \sin 2\theta_{e_{u}}
\sum_{\mu}M^{(\pi)}_{\mu}M^{(\sigma)}_{\mu}.
\end{equation}
Making use of the coefficients from Table 1 one might obtain general
expressions for matrix elements in the case of $\sigma$ states
$\gamma\mu=e_{u}x$ or $\gamma\mu=e_{u}y$:

$$
M^{(\sigma)}_{\mu}({\bf k}, {\bf{e}})=-\sqrt{6\pi}aB(k)e_{\mu} \kappa_{\mu}
\sin(\frac{ak_{\mu}}{2})+
$$
$$
\sqrt{\frac{8\pi}{3}} \{  (A_{0}(k)+A_{2}(k))e_{\mu} -
 3 A_{2}(k)\kappa_{\mu}( {\bf{e}} \cdot {\bf \kappa} )\}
\cos(\frac{ak_{\mu}}{2})
$$
\begin{equation}
 \qquad (\mu =x,y).
\end{equation}
Similarly,  one might obtain general expressions for matrix elements in the
case of $\pi$ states $\gamma\mu=e_{u}x$ and $\gamma\mu=e_{u}y$:
$$
M^{(\pi)}_{x}({\bf k}, {\bf{e}})=-\sqrt{6\pi}aB(k)e_{y} \kappa_{x}
\sin(\frac{ak_{y}}{2})+
$$
$$
\sqrt{\frac{8\pi}{3}} \{ (A_{0}(k)+A_{2}(k))e_{x} -
 3 A_{2}(k)\kappa_{x} ( {\bf{e}} \cdot {\bf \kappa} ) \}
\cos(\frac{ak_{y}}{2}),
$$

$$
M^{(\pi)}_{y}({\bf k}, {\bf{e}})=-\sqrt{6\pi}aB(k)e_{x} \kappa_{y}
\sin(\frac{ak_{x}}{2})+
$$
\begin{equation}
 \sqrt{\frac{8\pi}{3}} \{  (A_{0}(k)+A_{2}(k))e_{y} -
3 A_{2}(k)\kappa_{y} ( {\bf{e}} \cdot {\bf \kappa} ) \} \cos(\frac{ak_{x}}{2}).
\end{equation}



\vspace{0.5cm}

{\large \bf 4. Modelling the  polaronic matrix element effects in ARPES
spectroscopy} \vspace{0.2cm}

Orbital $b_{1g}-e_u$ quasidegeneracy and polaronic nature of the photohole
results in a complicated structure of the energy and momentum dependence of the
photointensity. Below, we would like to present some examples of the
straightforward model calculations of the matrix elements effects in ARPES.
First of all we should address to the ${\bf k}$-dependence of the single
$CuO_4$ center contribution, the polaron form-factor effects, the photon
polarization effects, and interference effects, caused by $b_{1g}-e_u$
quasidegeneracy.

\vspace{0.2cm}
 {\bf 4.1. Matrix element effects for isolated $CuO_4$ center}
\vspace{0.2cm}

The Figure 5 shows the contour-plots (the darker the color, the bigger the
photointensity) for the quantities $|M_{\gamma\mu}({\bf k})|^2$ which describe
a partial one-center form-factor contribution to the photo-current intensity.
The ${\bf k}$-dependence of the polarization averaged $\overline{
|\langle\Psi_{b_{1g}}({\bf{r}})|({\bf{e}}\cdot{\bf{r}})| e^{i{\bf
k}{\bf{r}}}\rangle|^{2} }$ for $\theta_{b_{1g}} = -0.33 \pi$ is shown in
Fig.5a. Simply speaking, this is a contribution of the conventional model
Zhang-Rice singlet. From left to right here we present the purely $Cu3d$,
purely $O2p$, and the total contributions, respectively. One should note the
complex ${\bf k}$-dependence of the oxygen contribution as compared with the
copper one.
 The ${\bf k}$-dependence of the polarization and orbitally averaged
$\overline{\sum_{\mu}|\langle\Psi_{e_{u}\mu}( {\bf{r}} )|({\bf{e}} \cdot
{\bf{r}})| e^{i{\bf k}{\bf{r}}}\rangle|^{2}}$ contribution of the purely oxygen
$e_u$ electron-removal state for $\theta_{e_{u}} = 0.30 \pi$ is shown in
Fig.5b. From left to right there are presented the $\pi$, $\sigma$, and hybrid
(interference) $\pi -\sigma$ contributions, respectively. It should be noted
that the $\pi$ and $\sigma$ partial contributions have a rather different ${\bf
k}$-dependence.

 To illustrate the photon polarization  effects we present
several examples of the angular (${\vec k}$) dependence of the photointensity
for the ''parallel'' (${\vec e}\parallel {\vec k}$), and ''perpendicular''
(${\vec e}\perp {\vec k}$) polarizations, respectively, calculated with the
help of Exps. (23)-(24). The Figure 6a relates to the $Cu3d$ partial
contribution to the photointensity with creation of photo-hole in ZR-singlet
state for the parallel ($\propto \cos ^{2} 2\phi $), or perpendicular ($\propto
\sin ^{2} 2\phi $) polarization, respectively. The Figure 6b relates to the
corresponding $O2p$ partial contribution. Here, one should note the more
complex form of the angular dependence due to the essentially different
structure of the respective matrix elements. However, the $k$ modulus
dependence of the polar plots in Fig.6b is rather weak. So, in a whole, the
polarization effects in both cases appear to be qualitatively, and even
quantitatively  similar. A comparative analysis of the Figures 5 and 6
indicates strong impact of the photon polarization effect on the final momentum
dependence of the photocurrent intensity.

\vspace{0.2cm}
 {\bf 4.2. Matrix element effects for isolated large polaron }
\vspace{0.2cm}

 To illustrate an important role of the matrix element effects,
we consider below a four-, and five-center model  of the immobile large PJT
polaron generated by the isolated valent $^{1}A_{1g}-{}^{1,3}E_{u}$ manifold of
the  $CuO^{5-}_{4}$ center,
 that is assuming a  localization of the photohole either in $b_{1g}$
or $e_u$ orbital on the  $CuO_4$ center.

The Figure 7a,b shows the contour-plots for a number of quantities
$|f^{(\gamma\mu)}_{\Gamma M}({\bf k})|^2$ which represent a peculiar
''$k$-portrait'' of the hole density within $N$-center large polaron ($N=4$,
Fig.7a, $N=5$, Fig.7b), and  describe a partial polaronic form-factor
contribution to the photo-current intensity. Again we see complex and various
momentum dependencies, reflecting both the hole symmetry and its distribution
in large polaron.

The Figure 8 shows the contour-plots for a number of quantities
$|{\cal{M}}_{\Gamma M}({\bf k})|^2$ ($\Gamma = B_{1g},E_u$) which describe a
polarization and orbital averaged overall matrix element effect in partial
$\Gamma$ contribution to the photo-current intensity with  one-center hole
basis consisting of the $b_{1g}$-, and $e_u$-orbitals. The top figures present
the total contribution, while below there are shown the partial $b_{1g}$-,
$e_u$-, and interference $b_{1g}-e_u$-contributions, respectively.
Interestingly, the interference term looks similarly in both cases. One should
notice the nonzero contribution of the $e_u$ states to the photointensity in
the BZ center ($\Gamma$-point) for the $E_u$ type polaron.

To illustrate the photon polarization  effects we have calculated the angular
(${\vec k}$) dependence of the  partial photointensity related to photo-hole
creation in $\Psi ^{e_{u}}_{B_{1g}}$ state  for the "parallel", and
"perpendicular" polarizations, respectively (Fig.9). Interestingly, the
dependence is qualitatively similar to the case of the ZR-singlet, at least for
the high-symmetry directions. In other words, the polarization dependence alone
could not distinguish the ZR-singlet among other terms with the relevant
symmetry. The Figures 8 and 9  convincingly illustrate the role played by the
matrix element effects, including the form-factor, orbital symmetry, photon
polarization, and interference effects. As we see, for a quantitative
comparison with experiment, it is necessary to add the ARPES amplitudes instead
of intensities, taking into account the polarization and energy of the incident
photons, and the direction of the photoemitted electrons.

So, one might unambiguously say that the matrix element effects result in a
complex ${\bf k}$ dependence of the photocurrent intensity which has to be
taken into account when addressing such issues as Fermi surface. Indeed, the
above model illustrations provide a wide choice of the "Fermi surface"-like
behavior.  Neglecting the matrix element effects results in erroneous
conclusions concerning the electronic structure of the electron-removal states.

\vspace{0.4cm} {\large \bf 5. Conclusions} \vspace{0.2cm}

We had not for an object the detailed fitting of the experimental photoemission
spectra, as we consider this problem in the meantime to be very complicated. At
present, there is no generally accepted model for the large PJT polaron in
copper oxides and appropriate ARPES spectral functions $A(\omega)$, and it
leads to uncertainties in quantitative interpretation of the experimental data.

Nevertheless, we see that the low-energy ARPES spectra could be originated from
the hole polaronic excitations which  complex spectral shape  is strongly
affected by the soft   lattice and  spin  fluctuations.  These usually have to
result in a complex ARPES spectral shape with a rather narrow purely electronic
($coherent$) peak and  structureless ($incoherent$) background which describes
the excitations accompanied by the emission and absorption of bosons (phonons,
spin waves ). One should notice that the polaronic spectral response  can
spread over a wide energy range of about  several tenths of eV.
 Experimental spectral   shape of the intensity is, qualitatively, compatible with that
  expected for spectral response of the PJT polaron \cite{ber}.
Perhaps, namely the polaron-like entity formation could explain the  extremely
narrow and intense peak lying below the Fermi energy and being the most
intriguing feature of ARPES in all the high-$T_c$ cuprates. We also could
propose that many unusual features of the cuprate ARPES including the ${\vec
k}$-dependence of spectral shape can be understood and described without any
Fermi-surface, large or small.

A qualitative comparison of experimental ARPES spectra and model calculations
of the matrix element effects allows to make a number of important conclusions:

1.  A model of the dispersionless valent $^{1}A_{1g}-{}^{1,3}E_{u}$ multiplet
of the two-hole $CuO^{5-}_{4}$ center and large ($N$=4, or 5) non-adiabatic PJT
polaron is quite enough compatible with experimental ARPES data. Moreover, this
is capable to describe some rather subtle spectral ARPES features, including
the ''remnant Fermi surface'' effect without any reference to itinerant
band-like states.
 It should be emphasized one more the principal role
of ${}^{1,3}E_{u}$ term determining the non-zero ARPES response in $\Gamma$
point.

2. The ${\bf k}$-dependent spectral weight transfer for the large PJT polaron
could be a natural origin of  a ''seeming'' hole dispersion when to be
described in a single-band model. At the same time, the polaronic transport
described by the Hubbard-like model also could provide the natural explanation
of the dispersion observed. Apparently, the straightforeward assignement of the
photocurrent intensity maxima to that of the quasiparticle spectral density may
result in erroneous conclusions.

3. Interpretation of the photoemission spectra for strongly correlated oxides
needs a caution and careful account for multi-band effects and matrix element
effects, especially in what concerns the ${\bf k}$-dependence of the spectral
weight, and Fermi surface assignement. Moreover, the polaronic nature of the
photo-hole gives rise to the problem of the ''third dimension'' of polaron, or
to its correlation length in $c$-direction resulting in a number of important
consequences concerning the photon energy dependence of photointensity.

All these conclusions cast doubt on results of numerous papers with simplified
interpretation of the low-energy ARPES data for $Sr_{2}CuO_{2}Cl_{2}$ made in
the framework of the various single-band versions of the  $t-J$ model
\cite{Wells,Nazarenko}, and aimed the interpretation of the ''experimental
quasiparticle dispersion law'' supposedly determined  from positions of  the
photocurrent intensity maxima. Unfortunately,  the available experimental data
do not allow to make so far the reliable conclusions on the nature of the
low-energy ARPES feature. The ARPES data have to be considered with great care
since it is most probable that they do not reflect straightforwardly the bulk
density of states (DOS) of the quasiparticle excitations. In our opinion, a
similar situation with the ARPES data interpretation occurs for many other
strongly correlated oxides, including the superconducting cuprates, that forces
to consider with caution many principal conclusions which are founded on the
ARPES data.

Concluding, one should be noted that the elaboration of adequate theory of the
electronic spectra for the strongly correlated oxides needs at present not only
solution of a number of complex theoretical problems, but the more perfect
experimental data with complete spectral, angular, and polarization analysis.

The research described in this publication was made possible in part by Award
No.REC-005 of the U.S. Civilian Research \& Development Foundation for the
Independent States of the Former Soviet Union (CRDF). The authors acknowledge a
partial support from the Russian Ministry of Education, grant \# 97-0-7.3-130.

\newpage

{\bf Figure captions}

Fig.1. Schematic representation of the electron (hole) density distribution in
the hybrid $Cu3dO2p$ $b_{1g}$-, and purely oxygen $e_{u}(\sigma ,\pi )$
molecular orbitals, assumed to be main components of the first electron-removal
state in insulating copper oxides.

Fig.2. Schematic structure of the large polaron: (a) $N = 4$, (b) $N = 5$.

Fig.3. Dependence of the $Cu3d$-atomic radial parameters $D_{1}(E)$ and
$D_{3}(E)$ on the photoelectron energy ($a_d=0.35$\AA).

Fig.4. Dependence of the $O2p$-atomic radial parameters $B(E)$, $A_{0}(E)$ and
$A_{2}(E)$ on the photoelectron energy ($a_p=0.52$\AA).

Fig.5.  Contour-plots for the ${\vec k}$-dependence of the partial one-center
form-factor contribution to the photo-current intensity for the depolarized
photons: a) $\gamma\mu = b_{1g}$, from left to right the $Cu3d$, $O2p$, and the
hybrid $Cu3dO2p$ ($\theta_{b_{1g}} = -0.3\pi$) contributions, respectively;
b)$\gamma\mu = e_{u}$, from left to right the $\pi$, $\sigma$ and hybrid
$\sigma \pi$ ($\theta_{e_{u}} = 0.4\pi$) contributions, respectively.

Fig.6. Photon polarization  effects. Angular (${\vec k}$) dependence of the
ZR-singlet partial contribution to photointensity for the "parallel" (${\vec
e}\parallel {\vec k}$), and "perpendicular"  (${\vec e}\perp {\vec k}$)
polarizations, respectively: a) $Cu3d$ partial contribution; b) $O2p$ partial
contribution.  Numbers near curves indicate the $k$ values.

Fig.7.  Contour-plots for polaronic formfactors $|f^{(\gamma\mu)}_{\Gamma
M}({\bf k})|^2$ which describe a partial polaronic form-factor contribution to
the photo-current intensity, and represent a peculiar ''$k$-portrait'' of the
hole density within $N$-center large polaron: a) $N=4$, b) $N=5$.

Fig.8. Contour-plots for  quantities $|{\cal{M}}_{\Gamma M}({\bf k})|^2$
($\Gamma = B_{1g},E_u$) which describe a polarization and orbital averaged
overall matrix element effect in partial $\Gamma$ contribution to the
photo-current intensity. The top figures present the total contribution, while
below there are shown the partial $b_{1g}$-, $e_u$-, and interference
$b_{1g}-e_u$-contributions, respectively.

Fig.9. Photon polarization  effects. Angular (${\vec k}$) dependence of the
partial contribution to photointensity related to photo-hole creation in $\Psi
^{e_{u}}_{B_{1g}}$ state for the "parallel", and "perpendicular" polarizations,
respectively. Different  curves correspond the same  $k$ values as in Fig.6b.

\newpage

\begin{table}
\caption{The coefficients $C^{(\gamma\mu)}_{m}({\bf{t}})$.}

\vspace{1.cm}
\begin{center}

\begin{tabular}{||c|c|c|c|c||} \hline

$C^{(\gamma\mu)}_{m}({\bf{t}})$ & $\frac{{\bf{x}}}{2}$ & $-\frac{{\bf{x}}}{2}$
& $\frac{{\bf{y}}}{2}$ & $-\frac{{\bf{y}}}{2}$
\\ \hline

$C^{(b_{1g})}_{+1}({\bf{t}})$ & $\frac{1}{2\sqrt{2}}$ &
$-\frac{1}{2\sqrt{2}}$ & $\frac{i}{2\sqrt{2}}$ &
$-\frac{i}{2\sqrt{2}}$ \\ \hline

$C^{(b_{1g})}_{-1}({\bf{t}})$ & $-\frac{1}{2\sqrt{2}}$ &
$\frac{1}{2\sqrt{2}}$ & $\frac{i}{2\sqrt{2}}$ &
$-\frac{i}{2\sqrt{2}}$ \\ \hline

$C^{(\sigma x)}_{+1}({\bf{t}})$ & $-\frac{1}{2}$ & $-\frac{1}{2}$
& $0$ & $0$ \\ \hline

$C^{(\sigma x)}_{-1}({\bf{t}})$ & $\frac{1}{2}$ & $\frac{1}{2}$ &
$0$ & $0$
\\ \hline

$C^{(\sigma y)}_{+1}({\bf{t}})$ & $0$ & $0$ & $\frac{i}{2}$ &
$\frac{i}{2}$
\\ \hline

$C^{(\sigma y)}_{-1}({\bf{t}})$ & $0$ & $0$ & $\frac{i}{2}$ &
$\frac{i}{2}$ \\ \hline

$C^{(\pi x)}_{+1}({\bf{t}})$ & $0$ & $0$ & $-\frac{1}{2}$ &
$-\frac{1}{2}$ \\ \hline

$C^{(\pi x)}_{-1}({\bf{t}})$ & $0$ & $0$ & $\frac{1}{2}$ &
$\frac{1}{2}$
\\ \hline

$C^{(\pi y)}_{+1}({\bf{t}})$ & $\frac{i}{2}$ & $\frac{i}{2}$ & $0$
& $0$
\\ \hline

$C^{(\pi y)}_{-1}({\bf{t}})$ & $\frac{i}{2}$ & $\frac{i}{2}$ & $0$
& $0$ \\ \hline
\end{tabular}
\end{center}

\vspace{1.cm}
\end{table}

\end{document}